\documentclass[twocolumn]{aastex631}

\graphicspath{{./}{figures/}}
\usepackage{amsmath}
\usepackage{threeparttable}
\usepackage{graphicx}        
\usepackage{CJK}
\usepackage{tablefootnote}

\newcommand\sref[1]{\hyperref[#1]{\S~\ref*{#1}}}
\newcommand\fref[1]{\hyperref[#1]{Fig.~\ref*{#1}}}
\newcommand\Eqref[1]{Eq.~(\hyperref[#1]{\ref*{#1}})}
\newcommand\eeqref[1]{Eq.~\hyperref[#1]{\ref*{#1}}}
\newcommand\tref[1]{\hyperref[#1]{Table~\ref*{#1}}}
\newcommand\aref[1]{\hyperref[#1]{Appendix~\ref*{#1}}}

\shorttitle{Where is the Super-Virial Gas?}
\shortauthors{Roy et al.}

\begin{document}
\title{Where is the Super-Virial Gas? The Supply from hot inflows}

\correspondingauthor{Manami Roy}
\email{roy.516@osu.edu}
\author[0000-0001-9567-8807]{Manami Roy}
\affiliation{Center for Cosmology and Astro Particle Physics (CCAPP), The Ohio State University, 191 W. Woodruff Avenue, Columbus, OH 43210, USA}
\affiliation{Department of Astronomy, The Ohio State University, 140 W. 18th Ave., Columbus, OH 43210, USA}
\author[0000-0003-1598-0083]{Kung-Yi Su}
\affiliation{Black Hole Initiative, Harvard University, 20 Garden St., Cambridge, MA 02138, USA}
\author[0000-0002-4822-3559]{Smita Mathur}
\affiliation{Center for Cosmology and Astro Particle Physics (CCAPP), The Ohio State University, 191 W. Woodruff Avenue, Columbus, OH 43210, USA}
\affiliation{Department of Astronomy, The Ohio State University, 140 W. 18th Ave., Columbus, OH 43210, USA}
\author[0000-0002-7541-9565
]{Jonathan Stern}
\affiliation{School of Physics \& Astronomy, Tel Aviv University, Tel Aviv 69978, Israel}

\begin{abstract} \label{abstract}
To understand the presence of the super-virial temperature gas detected in the Milky Way, we present our findings from isolated galaxy simulations of {Milky Way-mass} systems using {\small GIZMO} with the FIRE-2 (Feedback In Realistic Environments) stellar feedback model. It unveils the presence of a significant super-virial temperature {($T>T_{vir}$)} gas component within 20 kpc from the galactic center. {We also find that 70--90\% of the total super-virial gas is extra-planar, at $1<z<6$ kpc and $R_{\rm cyl}<15$ kpc.} This super-virial gas has a mass of $1-2\times10^7$ ${\rm M_\odot}$ {with typical gas densities are $10^{-3.5}-10^{-2.5} \, \rm cm^{-3}$}. We find that some of the virial gas ($T\sim10^6$K) forms a rotating hot inflow, where gravitational energy is converted to thermal energy mainly via compressive heating. This process causes gas falling close to the rotation axis to reach super-virial temperatures {via a combination of compressive heating and {shocks}} just before cooling and joining the disk. Stellar feedback heating accounts for less than 1\% of the super-virial gas, indicating its minimal influence despite expectations. Even in scenarios with no stellar feedback effects considered, abundant super-virial gas persists, highlighting the dominance of alternative heating mechanisms. We also show that cosmic rays do not have a significant effect on heating the gas to a super-virial temperature. Our study illuminates the intricate dynamics of hot virial and super-virial gas surrounding {Milky Way-mass} galaxies, emphasizing the prominent role of infall-driven compressive {and shock-heating} processes in shaping thermal evolution. 
 \end{abstract}
\keywords{Galaxy: evolution --- Galaxy: halo --- methods: numerical --- (ISM:) cosmic rays --- (galaxies:) quasars: absorption lines --- X-rays: galaxies}

\section{Introduction}
\label{S:intro}

Galaxies are the building blocks of our universe, and how galaxies form and evolve is an active area of research. A spiral galaxy has two main components: a star-forming disk and a diffuse gaseous halo surrounding the disk. This diffuse gaseous halo, known as the Circumgalactic medium (CGM) \citep{Tumlinson2017, 2023ARA&A..61..131F}, plays a crucial part in galaxy evolution. The CGM is the habitat for large-scale gas flows that fundamentally play an important role in the evolution of the galaxy by providing fresh and recycled gaseous fuel for star formation and regulating the galaxy's interactions with other galaxies. A comprehensive understanding of the CGM stands to illuminate key unresolved questions about the processes driving galaxy formation and evolution.

Unfortunately, CGM is very hard to detect in emission due to its low density. However, from absorption lines in the spectra of background quasars, we can infer the physical properties of the CGM.  When light from a background quasar passes through the CGM of a galaxy, it suffers absorption from the metals and ions in the CGM. Absorption studies made by COS spectrograph onboard HST along with XMM-Newton and Chandra have enabled us to detect this multiphase diffuse gas around {Milky Way (MW)-like} galaxy in UV [e.g. \cite{Tumlinson2011, Werk2016}] and X-ray [e.g. \cite{Gupta2012, Fang2015, 2021ApJ...918...83D, Mathur2022}] respectively. The MW CGM has also been detected in X-ray emission [e.g., \cite{Snowden2000, HenleyShelton2013, Naka2018}].

Until 2019, these observations led us to a picture where the CGM is multi-phase in nature and has three main components; virial temperature warm-hot gas (T$\sim 10^{6-6.5}$ K), warm gas (T$\sim 10^{5-5.5}$ K) and cool gas (T$<10^4$ K). However, this picture got modified with the recent discovery of the super-virial component (T$\sim 6\times10^{6} \rm K -2\times10^{7}$K; \cite{2019ApJ...882L..23D}) in the MW CGM using absorption lines from XMM-Newton observation towards the Blazar IES 1553+113. 
Since then, there have been multiple detections of this hotter super-virial gas in both emission and absorption \citep{2019ApJ...887..257D, Gupta2023, 
 2023ApJ...952...41B, 2023ApJ...946...55L, 
 2024MNRAS.527.5093M}. In emission, \cite{2019ApJ...887..257D} detected the super-virial component towards the same sight-line of Blazar IES 1553+113 using XMM-Newton, and \cite{Gupta2021} found this component along with the virial component toward four sightlines from Suzaku and Chandra. Later, \cite{2023ApJ...952...41B} also detected this super-virial gas in the MW CGM in emission toward the Blazar Mrk 421 and 5 other sightlines near the Blazar field.  Even a recent study by \cite{2023ApJ...946...55L} detected SiXIV K$\alpha$ and SXVI K$\alpha$, associated with a gas phase with super-virial temperature, in the CGM of the MW, with the use of stacked X-ray spectral observations towards multiple different extragalactic sightlines.

The super-virial component has not only been detected towards some particular sightlines, but all-sky surveys from {Suzaku}, Halo-Sat, and eROSITA have also found this component abundantly all over the sky. {\cite{Gupta2023} found this component abundantly while analyzing the X-ray emission of the shells of the eROSITA bubbles from Suzaku data.} \cite{Bluem} utilized observations from the HaloSat all-sky survey to delve into the emissions from the MW's halo and detected the super-virial gas. \cite{Ponti}, in their study of the MW halo within the eFEDDS field employing eROSITA, also detected this component.  

This discovery begs questions like: \textit{where is this gas situated; in the ISM, inner CGM, or outer CGM? How is this gas formed: through feedback or some other heating mechanisms like cosmic ray (CR) heating? What is the nature of this gas: is it as diffuse as the volume-filling virial temperature gas, or is it more or less dense? }


In this paper, we will address the above questions in light of an idealized simulation. We investigate a suite of simulations of {MW-mass host galaxy}, where we focus on {this super-virial gas component}. The paper is structured as follows. In Section \ref{S:methods}, we discuss the methodology of our simulation, where we describe the initial conditions of our simulation setup (in Section \ref{S:in}) along with our definition of super-virial CGM gas for our analysis (in Section \ref{S:def_sup}). In Section \ref{S:results}, we demonstrate our results from our analysis of the simulations, where we discuss the location (Section \ref{S:location}), density (in Section \ref{S:den}), amount (Section \ref{S:amount}), and origin (Section \ref{S:origin}) of the super-virial gas in the CGM. We also discuss if feedback and cosmic rays heating have any effect on the origin of this gas in Section \ref{S:location} and \ref{S:amount}. We discuss the limitations of our work and planned future work in Section \ref{dis}. Finally, we summarize our results and discuss future work in Section \ref{S:conclusion}. 


\section{Methodology} \label{S:methods}
Our simulations utilize {\small GIZMO}\footnote{A public version of this code is available at \href{http://www.tapir.caltech.edu/~phopkins/Site/GIZMO.html}{\textit{http://www.tapir.caltech.edu/$\sim$phopkins/Site/GIZMO.html}}}  \citep{2015MNRAS.450...53H}, in its meshless finite mass (MFM) mode, which is a Lagrangian mesh-free Godunov method, capturing the advantages of grid-based and smoothed-particle hydrodynamics (SPH) methods. Numerical implementation details and extensive tests are presented in a series of methods papers for, e.g.,\ hydrodynamics and self-gravity \citep{2015MNRAS.450...53H}, magnetohydrodynamics \citep[MHD;][]{2016MNRAS.455...51H,2015arXiv150907877H}, anisotropic conduction and viscosity \citep{2017MNRAS.466.3387H,2017MNRAS.471..144S}, and cosmic rays \citep{chan:2018.cosmicray.fire.gammaray}. 

In all simulations except for the runs with no feedback, we incorporate the FIRE-2 model, an implementation of the Feedback In Realistic Environments (FIRE) approach. This encompasses treatments of the interstellar medium (ISM), star formation, and stellar feedback, the details of which are given in \citet{hopkins:sne.methods,2017arXiv170206148H} along with extensive numerical tests. 
Our simulations cover a broad temperature range from $10$ to $10^{10}$ K, accounting for various cooling mechanisms, including photoelectric and photo-ionization heating, collisional, Compton, fine structure, recombination, atomic, and molecular cooling.

Star formation is facilitated through a sink particle approach, allowed only in molecular gas regions with densities surpassing $100{\rm cm^{-3}}$, where self-shielding and local self-gravitation dominate. Upon formation, star particles are treated as cohesive stellar populations, inheriting the metallicity of their parent gas particles. Feedback mechanisms, including supernovae and mass-loss rates, are derived from IMF-averaged values calculated from {\small STARBURST99} \citep{1999ApJS..123....3L} with a Kroupa initial mass function (IMF) \citep{2002Sci...295...82K}.

The stellar feedback model encompasses diverse processes: (1) Radiative feedback, involving photo-ionization, photo-electric heating, and single and multiple-scattering radiation pressure tracked in five bands  (ionizing, FUV, NUV, optical-NIR, IR), (2) Continuous stellar mass loss and injection of mass, metals, energy, and momentum through OB and AGB winds, and (3) Type II and Ia supernovae (including both prompt and delayed populations), injecting appropriate mass, metals, momentum, and energy into the surrounding gas based on tabulated rates.
All the simulations integrate magnetohydrodynamics (MHD), fully anisotropic conduction, and viscosity, employing the Spitzer-Braginski coefficients to capture the relevant physics.

The model of cosmic ray (CR) treatment includes streaming, which occurs at the local Alfv\'en speed or sound speed, whichever is larger, incorporating an appropriate streaming loss term that thermalizes energy, as described in \cite{Uhlig2012}. Diffusion is modeled with a fixed diffusivity $\sim10^{29}\, \rm cm^2 s^{-1}$ alongside adiabatic energy exchange between gas and CR pressure and includes hadronic and Coulomb losses (following \cite{Guo2008}). We consider a single energy bin for GeV CRs, which dominate the pressure, and treat them in the ultra-relativistic limit. Both streaming and diffusion are fully anisotropic along magnetic field lines. CRs are injected by supernovae (SNe), with 10\% of each SNe's energy being transferred into CRs, consistent with studies like \cite{Pfrommer2017a, Pfrommer2017b}. Details on cosmic ray physics are demonstrated in \cite{Su2019, Chan2019}.

\subsection{Initial conditions} \label{S:in}
The initial setup follows the detailed specifications outlined in \cite{Su2019, Roy23}. To further stabilize the host CGM, the simulation region is expanded to 3 times the viral radius, and the simulations are run adiabatically (no cooling or star formation) for 4.5 Gyr to relax any initial transients before the satellites are placed into the CGM. 
The simulation properties are summarized in Table 1 of \cite{Roy23}. In this paper, we only focus on the {\bf m12} halo which is corresponding to the {MW-mass} halo of mass $\sim1.8\times10^{12}\,{\rm M_{\odot}}$.


We consider only the runs with 20 satellites of $2\times10^{9}\,{\rm M_{\odot}}$ DM halo mass ({\bf 20xm09}). 

We vary different physical properties to test the impact of stellar feedback and cosmic rays.  
{Our runs are the following :
\begin{itemize}
\item \textbf{m12\_lr:} low resolution ($8e4\,M_\odot$) and without CR
\item \textbf{m12\_hr:} high resolution ($8e3\,M_\odot$) and without CR
\item \textbf{m12\_lr\_nofb:} low resolution ($8e4\,M_\odot$), without CR and feedback. {It is important to note that a galaxy without feedback can not represent a realistic system; it is therefore not intended to match observed galaxies, but rather serves as a baseline to isolate and highlight the role of feedback.} 
\item \textbf{m12\_lr\_CR\_low:} low resolution ($8e4\,M_\odot$) and with low CR
\item \textbf{m12\_lr\_CR\_high:} low resolution ($8e4\,M_\odot$) and with high CR
\end{itemize}}

We initialize the dark matter (DM) halo, bulge, black hole, and gas+stellar disk of m12 by following \cite{1999MNRAS.307..162S,2000MNRAS.312..859S}.
The initial metallicity goes down from solar ($Z=0.02$) to $Z=0.001$ with radius as $Z=0.02\,(0.05+0.95/(1+(r/20\,{\rm kpc})^{1.5}))$. Magnetic fields are initialized azimuthal with $|{\bf B}|=0.03\,\mu{\rm G}/(1+(r/20\,{\rm kpc})^{0.375})$. More details of the initial setup are described in the Methodology section of \cite{Roy23}. 
{We initialize CRs in the high CR runs such that the CR energy density is in equipartition with the thermal energy density. In this case, we decrease the temperature by half, keeping the density the same, and put the rest of the energy in CR.
In low CR runs, CRs are initialized such that the CR energy density is in equipartition with magnetic energy density, which is $\sim5$ order of magnitude lower than the thermal energy density.} 

\begin{figure*}
\includegraphics[width=0.33\textwidth]{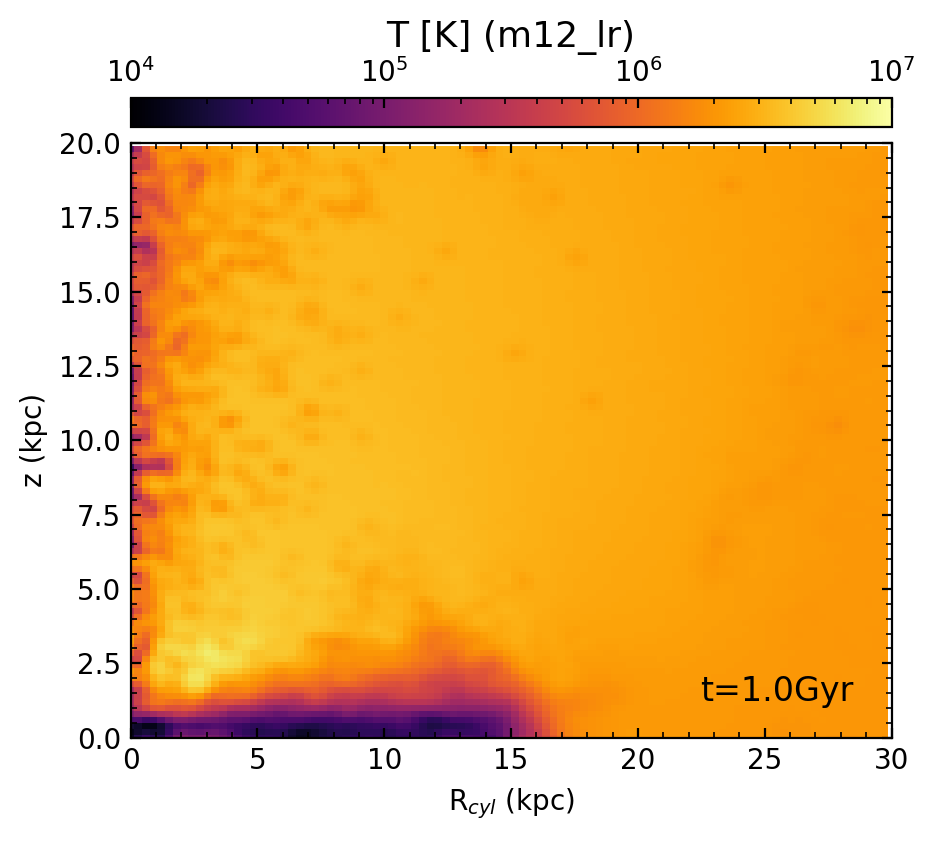}
\includegraphics[width=0.33\textwidth]{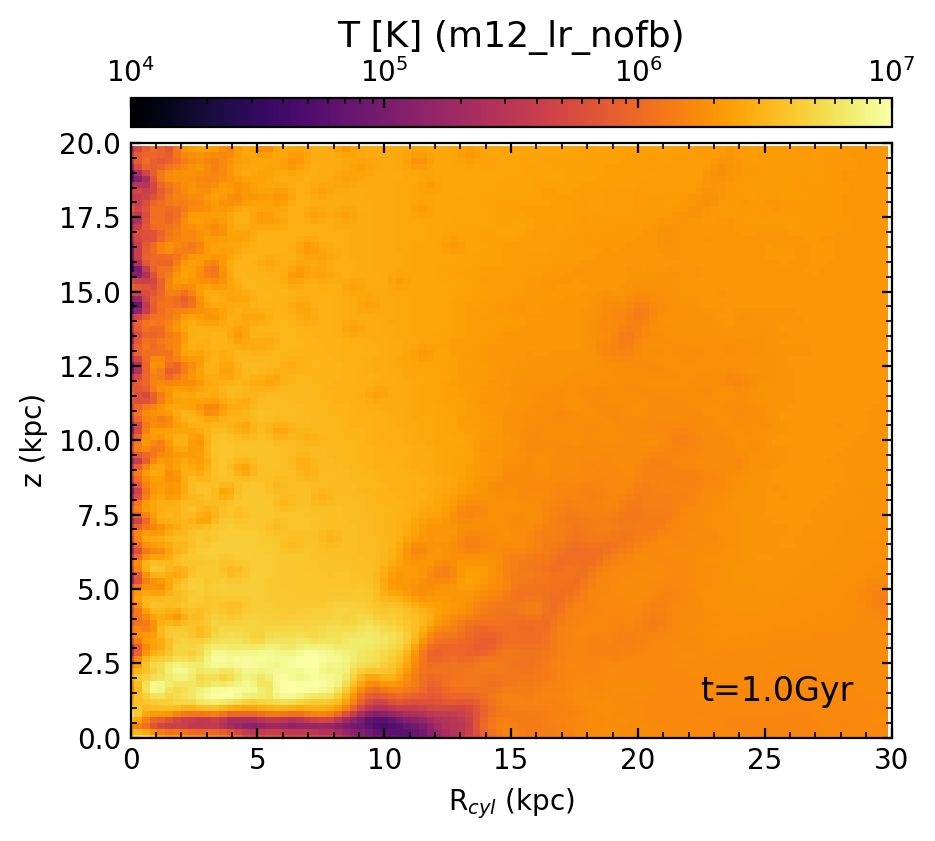}
\includegraphics[width=0.33\textwidth]{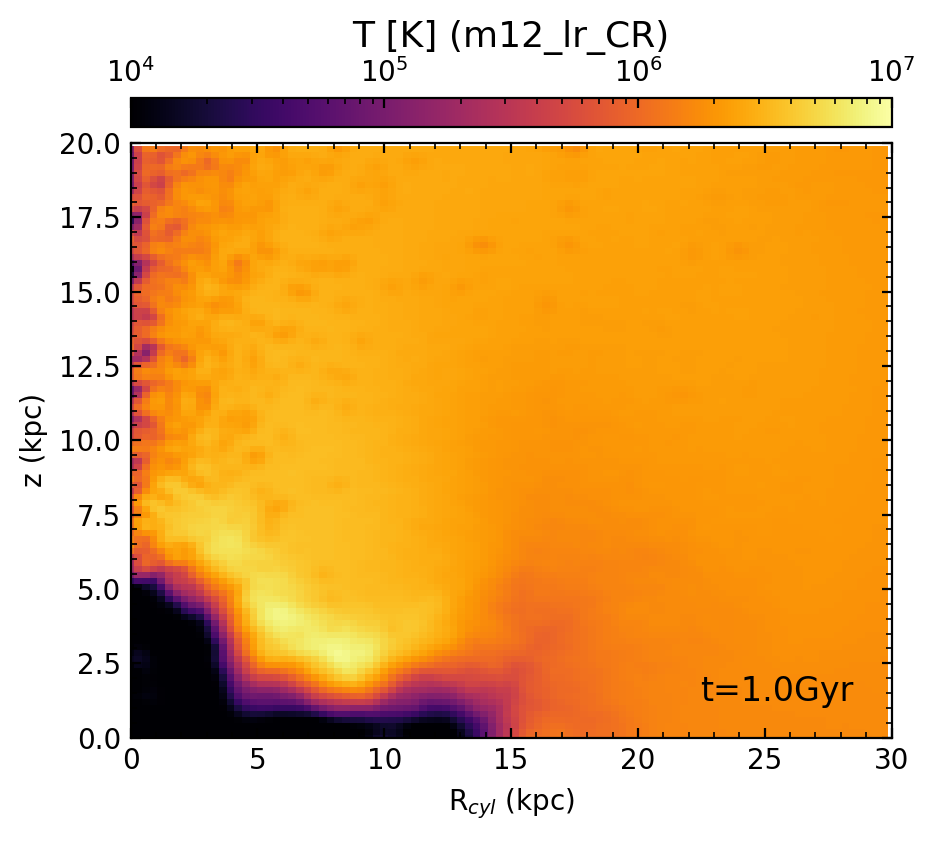}

\includegraphics[width=0.33\textwidth]{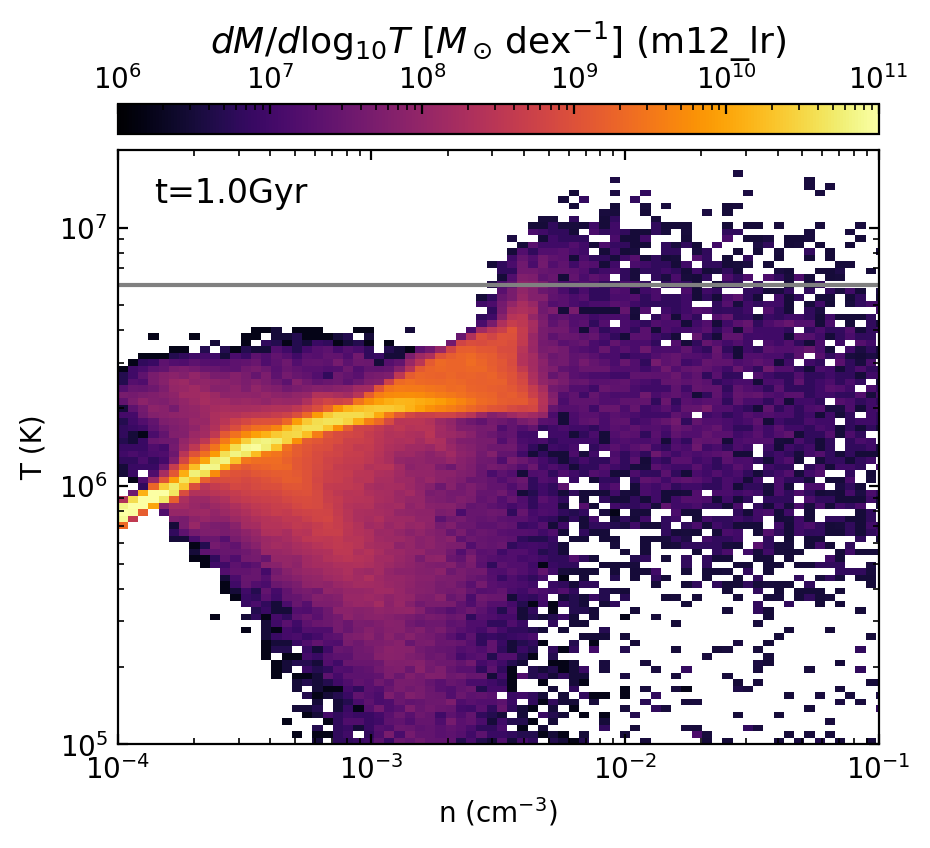}
\includegraphics[width=0.33\textwidth]{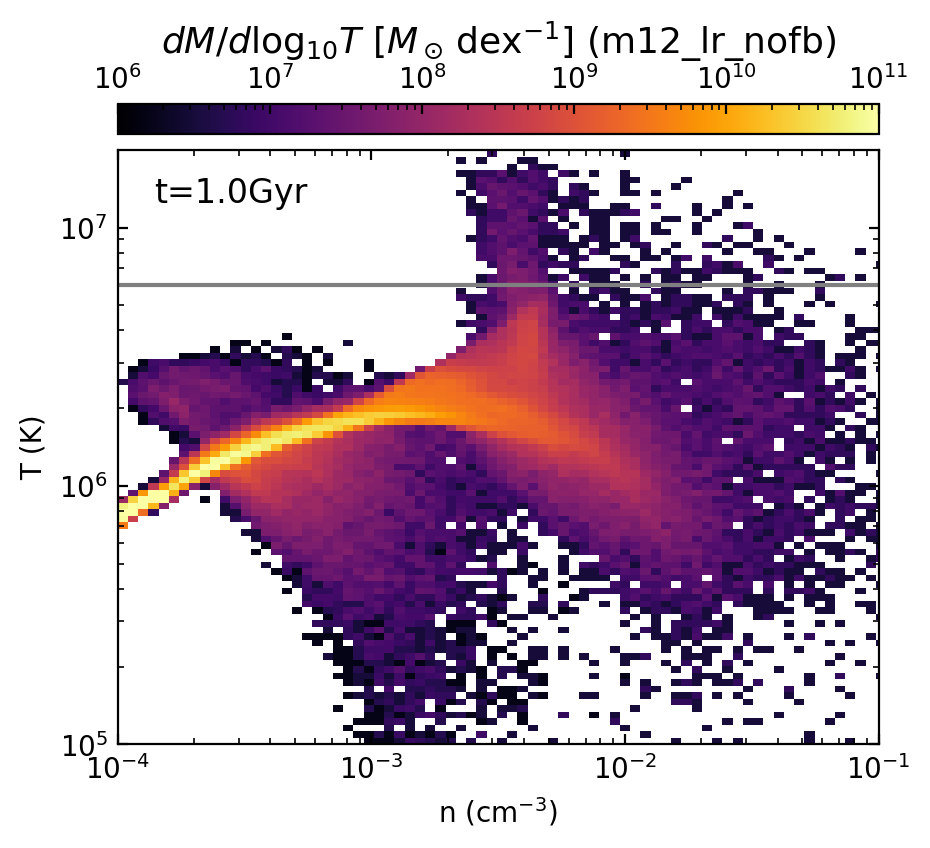}
\includegraphics[width=0.33\textwidth]{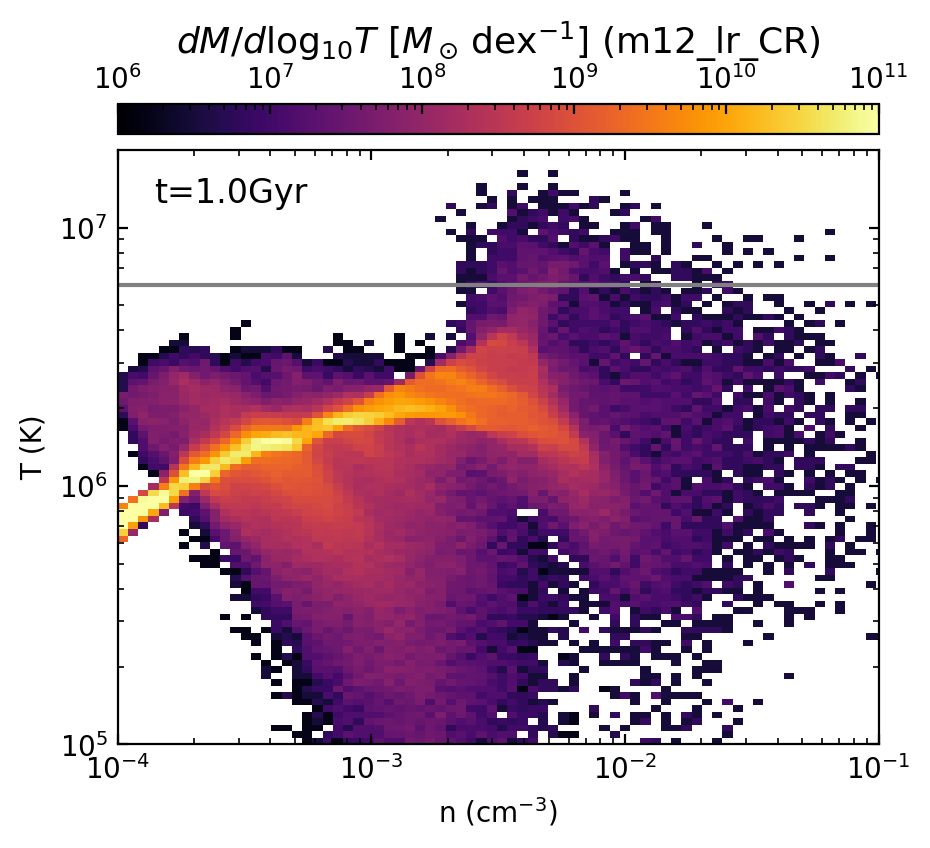}
\caption{{\textit{Top:} Top row denotes the gas temperature distribution of gas in the $R-z$ plane, which shows that super-virial gas ($T>6\cdot 10^{6}$ K) is mostly in the extra-planar region of the galaxy at $R<15$ kpc and $1<z<6$ kpc. \textit{Bottom:} Temperature-density distribution. The density of the super virial gas phase is $10^{-3.5}-10^{-2}$ cm$^{-3}$}.}
\label{f:2d-pdf}
\end{figure*}

We initialize the satellites similarly, like the host described above, though without a CGM gas halo. {The satellites are initialized orbiting randomly between 100 kpc to 150 kpc of the host galaxy in a circular orbit with the circular velocity of the host galaxy at that radius.} The properties are also summarized in Table 1 of \cite{Roy23}. {The satellite configuration was originally designed to see satellite contribution to CGM properties, and we find that they have relatively little contribution in producing the super-virial phase. Therefore, we focus mainly on the host gas for this work.}

\subsubsection{Gas temperature of the initial condition}
All the runs have similar initial setups, except for the high CR run. We check that our initial setups for different runs do not contain any super-virial gas. We confirm that the initial temperature is not more than $4\times10^6$K in all the runs. In the high CR run, the initial temperature profile is even lower for maintaining hydrostatic pressure balance, not crossing $2\times10^6$K. The variation in the initial condition will be useful to test whether the super-virial gas can be found even with a cooler halo, and test whether the results are sensitive to the initial setup.  We also confirm that the initial density ranges similarly in all the cases, with virial gas having density $10^{-(3-4)}$ $\rm cm ^{-3}$.
\subsection{Defining Super-virial Gas} \label{S:def_sup}
{Observationally, the gas with a temperature 
greater than the virial temperature $\sim2-3\times10^6$ has been identified as the super-virial phase. For this analysis, we use the temperature cut of  
$\rm T>6\times10^6$K for the super-virial temperature gas. Thus, we consider gas with a temperature above this threshold as the `super-virial' gas.}  



\section{Results}
\label{S:results}
\subsection{Where is the super-virial gas?}
\label{S:location}
This section delves into the investigation of the distribution of super-virial gas within the simulation box. {We present the temperature distribution of gas in the cylindrical radius (R$_{\rm cyl}$) and galactic height (z) plane, in the top row of Figure \ref{f:2d-pdf}. The distribution denotes the three different runs at a 1 Gyr snapshot of the simulation for no CR (1st column), no CR and no Feedback (2nd column), and the low CR (3rd column) cases.} In all the 2-D distribution plots, we did a radial cut of 6 times the gas scale radius region around the satellites to exclude the ISM of the satellites, as we chose this radius of the satellite to be the radius beyond which the gravitational pull from the satellites is negligible. Note, however, that we do not exclude satellite particles that can be stripped from the satellites. 

{The top panel shows that the super-virial phase is in the extra-planar region in all the cases, at $1<\lvert z \rvert<6$ and $R_{\rm cyl}<15$ kpc. Note that 6 kpc is the gas disc exponential scale length of the galaxy, and 1 kpc is the stellar bulge
Hernquist-profile scale length. Thus, we see that the super-virial gas mainly extends up to the disc gas scale length but beyond the stellar bulge scale length for a galaxy. However, the different physical processes lead to different extra-planar geometries. With no feedback, the super-virial gas is in a flat geometry and distributed more in R$_{\rm cyl}$. With stellar feedback, the gas is more concentrated at inner R$_{\rm cyl}$ and inner z. With CR feedback, the gas is more extended in z.}

\begin{figure*}
\includegraphics[width=0.49\textwidth]{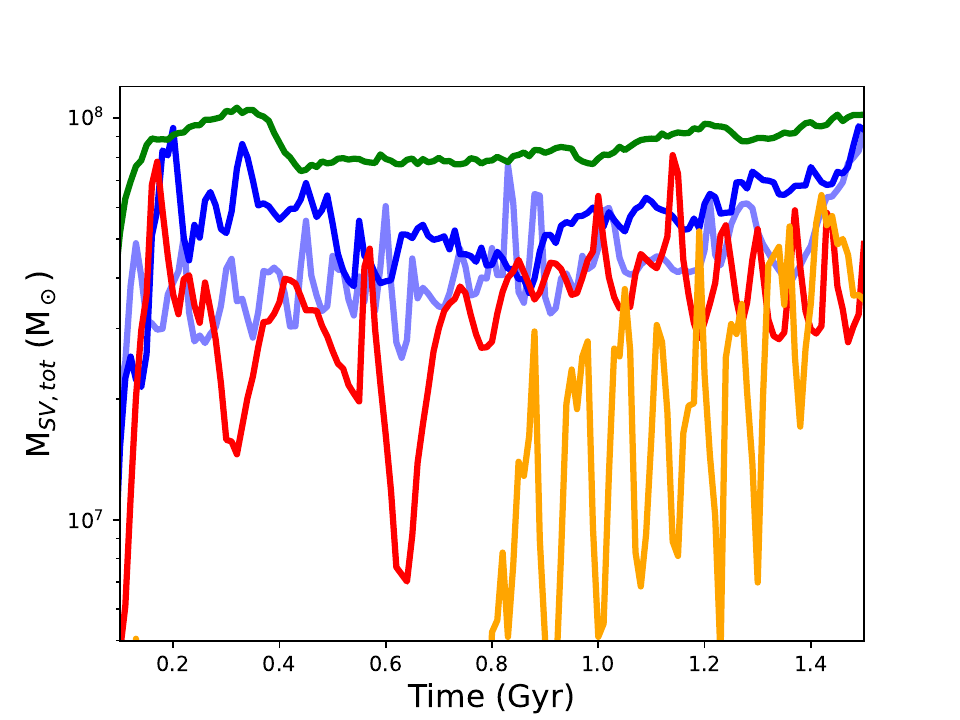}
\includegraphics[width=0.49\textwidth]{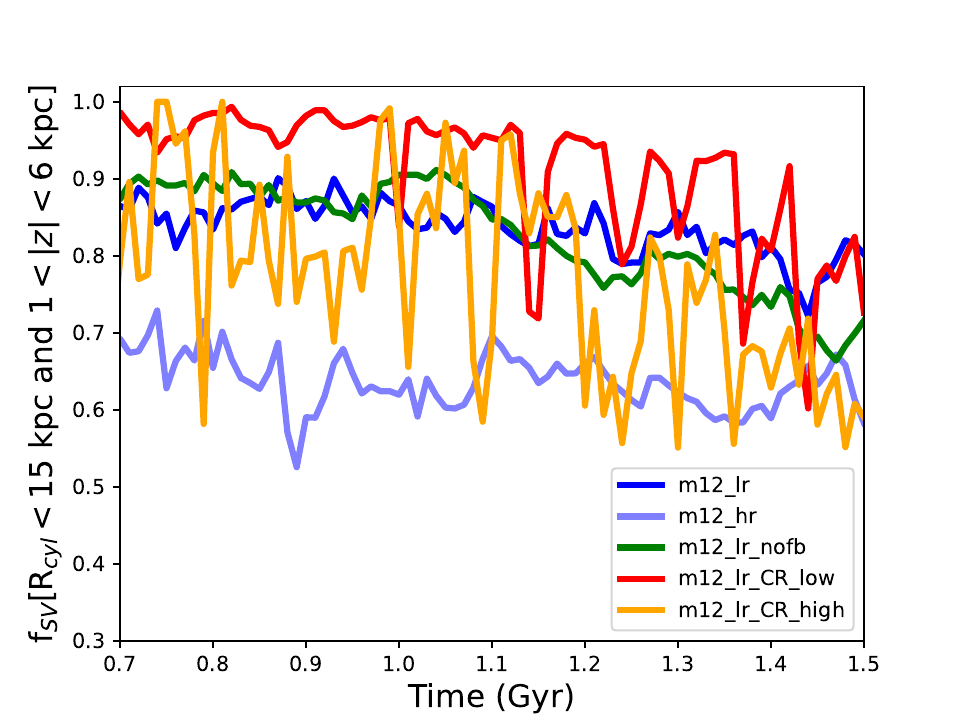}
\includegraphics[width=0.49\textwidth]{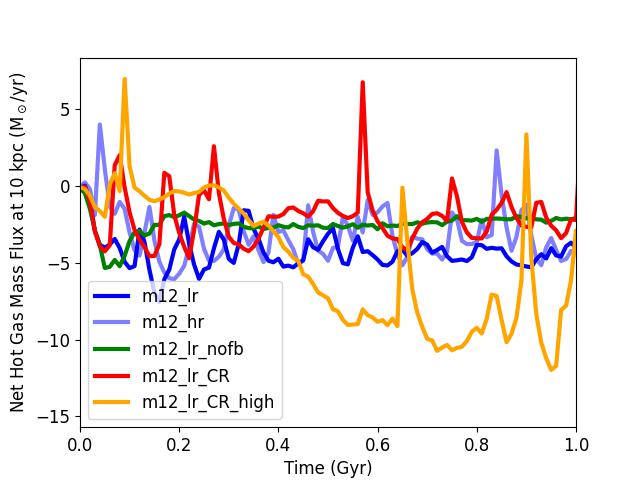}
\includegraphics[width=0.49\textwidth]{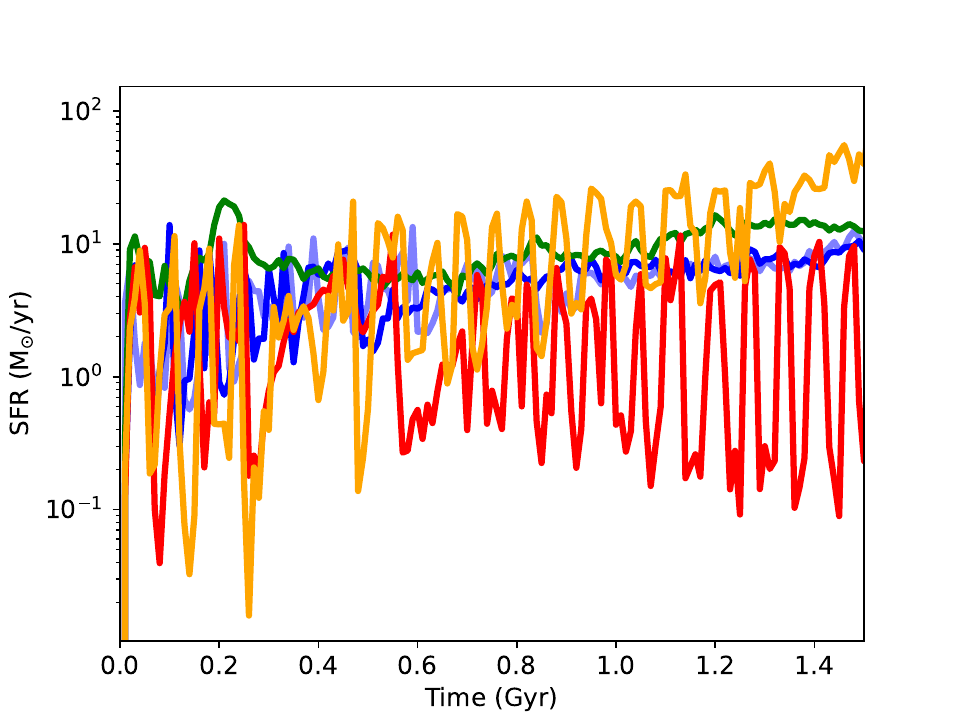}
\caption{{\textit{Top Left:} Time evolution of the total super-virial gas (T $>6\times10^6$ K) mass in various simulation runs. The figure demonstrates the persistent presence of super-virial gas across different runs, including those without stellar feedback. Notably, substantial amounts of super-virial gas are retained even in simulations with cosmic rays and varying initial conditions, such as the high cosmic ray run. \textit{Top Right:} Fraction of total super-virial gas within $R_{\rm cyl}<15$ kpc and $1<\lvert z \rvert<6$ kpc (6 kpc is the exponential scale-length of the gaseous disk). \textit{Bottom Left:} Time evolution of the mass flux of hot inflow at 10 kpc. \textit{Bottom Right:} Time evolution of the star formation rate of the host galaxy.}}
\label{f:time_evo_sv}
\end{figure*} 

\subsection{How dense is the super-virial gas?}
\label{S:den}

In this section, we scrutinize the density distribution of the super-virial gas. {In the bottom panel of Figure \ref{f:2d-pdf}, we show the temperature density 2-d distribution weighted by mass. The horizontal line {in the bottom panel of} Figure \ref{f:2d-pdf} denotes the super-virial temperature cut of $6\times10^6$K; gas with the temperature above this threshold is called `super-virial'. The temperature-density phase-space of the gas shows that the density of the super-virial gas is roughly in between $10^{-3.5}-10^{-2.5}$ cm$^{-3}$. This density is higher than the virial volume filling gas density as the gas is closer to the disk, as shown in the top panel of Figure \ref{f:2d-pdf}. Note that the no-feedback run has a pillar of lower-density gas in the super-virial phase, arising from the more distributed geometry mentioned in the above section.}



\subsection{How much super-virial gas is there in the galaxy?}
\label{S:amount}
Next, we quantify the mass of super-virial gas present in the galaxy. {We plot the total mass of the super-virial gas in the top-left panel of Figure \ref{f:time_evo_sv} for all the runs.
Given that the bulk of the super-virial gas resides within  R$_{cyl}<15$ kpc and $1<\lvert z \rvert<6$ kpc, we plot the fraction of total super-virial gas present within this region in the top-right panel of Figure \ref{f:time_evo_sv}.} 
{Here, we do not exclude the region around the satellites, as we would like to check if satellites also contribute to this gas phase. As the satellites are in circular orbits between 100 kpc and 150 kpc, the regions around them are automatically cut out while we calculate the super-virial gas mass inside 20 kpc and 10 kpc.} Key insights from this analysis include:
\begin{itemize}
\item \textbf{Mass:} The mass of {total super-virial gas fluctuates within the range of $0.2-1\times10^8$ ${\rm M_\odot}$ throughout the entire 1.5 Gyr simulation period.} CGM gas mass being $2\times10^{11}{\rm  M_\odot}$, the super-virial gas is only 0.1-1\% of the total CGM gas, hence sub-dominant in the baryonic budget and does not contribute to the missing baryon problem. {The top-right panel of Figure \ref{f:time_evo_sv}  shows that 70--90\% of total super-virial gas is within $R_{\rm cyl}<15$ kpc and $1<\lvert z \rvert<6$ kpc, again supporting the extra-planar geometry.}

\item \textbf{Effect of Feedback:} The presence of the super-virial gas could have been a direct result of heating by stellar feedback. Therefore, if we do not include feedback in our simulations, we might expect no (or a significantly smaller amount of) super-virial gas. Surprisingly, however, the simulations with no feedback display a higher amount of mass of super-virial gas among all the runs. It also indicates that in our simulations, the stellar feedback is not predominantly responsible for generating the super-virial gas. 

\item \textbf{Effect of Cosmic Rays:} CR heating by streaming heating or Coulomb heating can be one of the heating sources of the super-virial gas. CR can also drive outflows that can heat the gas. We investigate these effects with two runs with CRs, one with low CR and one with high CR. {The mass of super-virial gas exhibits greater temporal fluctuations with increasing cosmic ray (CR) feedback, although the overall magnitude of the mass budget remains qualitatively similar. This suggests that while CR feedback induces significant variability in the super-virial gas content, it does not directly contribute to its origin.}  

\item \textbf{Effect of Resolution:} 
We have conducted higher-resolution simulations (10 times higher resolution compared to our low-resolution run), represented by {light-shaded blue line} in Figure \ref{f:time_evo_sv}. We anticipate that these high-resolution (hr) runs would capture denser, super-virial gas than the low-resolution (lr) counterparts. {In the top-left panel of Figure \ref{f:time_evo_sv}, we observe that the total super-virial gas mass in the hr runs qualitatively similar to that in the lr runs. 
The top-right panel of Figure \ref{f:time_evo_sv} shows the amount of gas within a 15 kpc radius and
1–6 kpc height. Here, we see a 20\% difference between the high- and low-resolution simulations. 
The remaining 20\% super-virial gas in the high-resolution run is present at $z>6$ kpc. This is because the denser and hotter structures at higher galactic heights are resolved by the high-resolution run. {While these results demonstrate improved capture of vertical structure in the high-resolution run, we are currently limited by computational constraints and can not explore even higher resolutions. Overcoming this limitation is an important direction for future work.} 


\item \textbf{Effect of Other Satellite Runs:} We also check the other satellite runs described in \cite{Roy23}, with m10 and m08 satellites, to check whether the satellite distribution has a significant effect on our results (not shown in the figure). We find that there is no significant difference in the super-virial gas mass budget with the difference in satellite distribution. {However, we discuss how our result can depend on the satellites' initial conditions in the discussion section.}}
\end{itemize}
\begin{figure*}[h!]
\includegraphics[width=0.98\linewidth,height=0.76\paperheight]{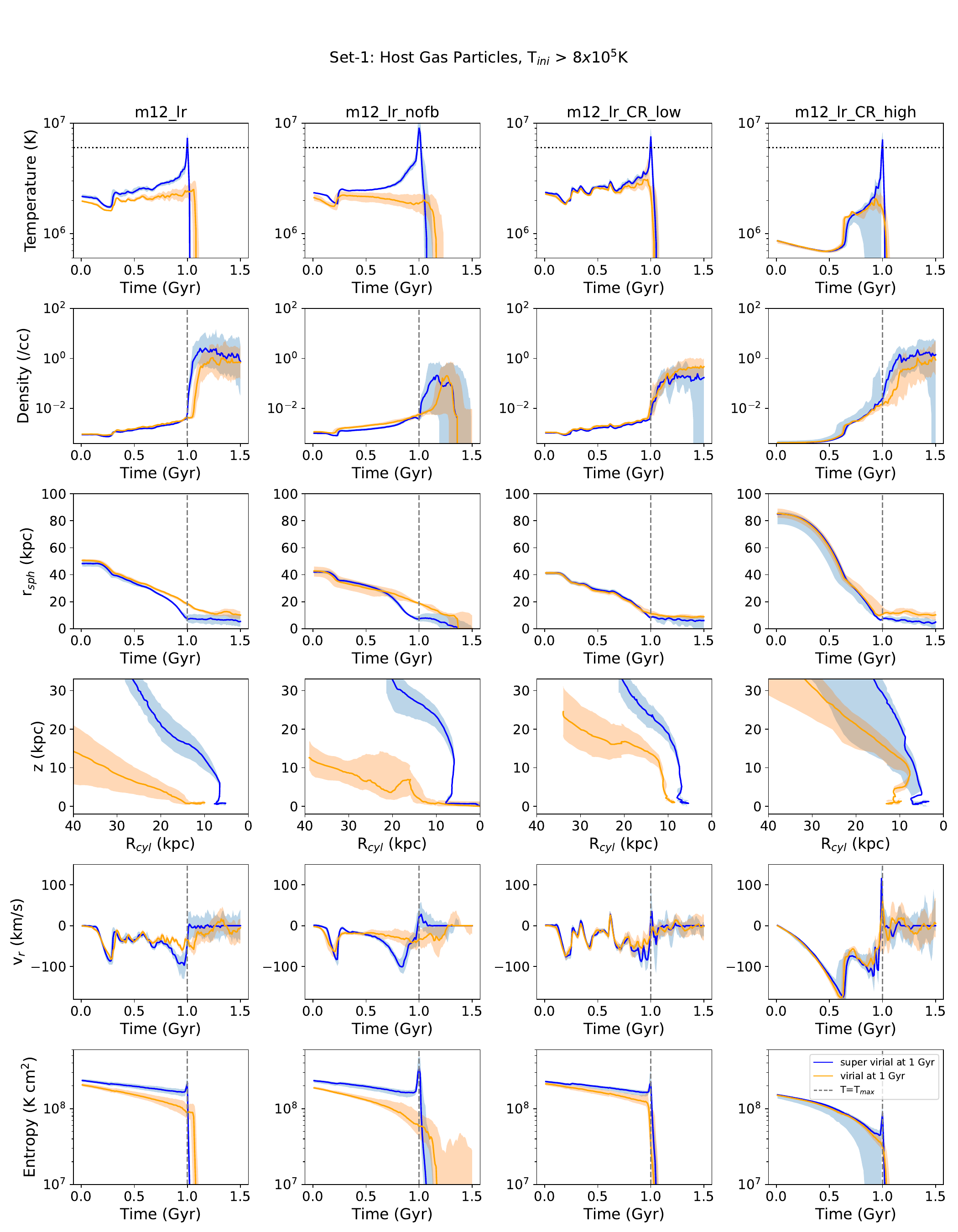}
\caption{{Time evolution (solid line: median and shaded region: percentile error) of the temperature (top row), density (2nd row), spherical radius (3rd row), radial velocity (5th row) and entropy (6th row) of the host gas particles with initial temperature T$_{\rm ini} \sim 2 \times 10^6$ K, which reached their maximum temperature at a simulation time of 1 Gyr. The 4th row denotes the particle trajectories in the $R_{\rm cyl}-z$ plane.} Blue and orange indicate gas with super-virial and virial temperatures at 1 Gyr, respectively.} 
\label{f:temp_sv}
\end{figure*}

\begin{figure}
\includegraphics[width=\linewidth]{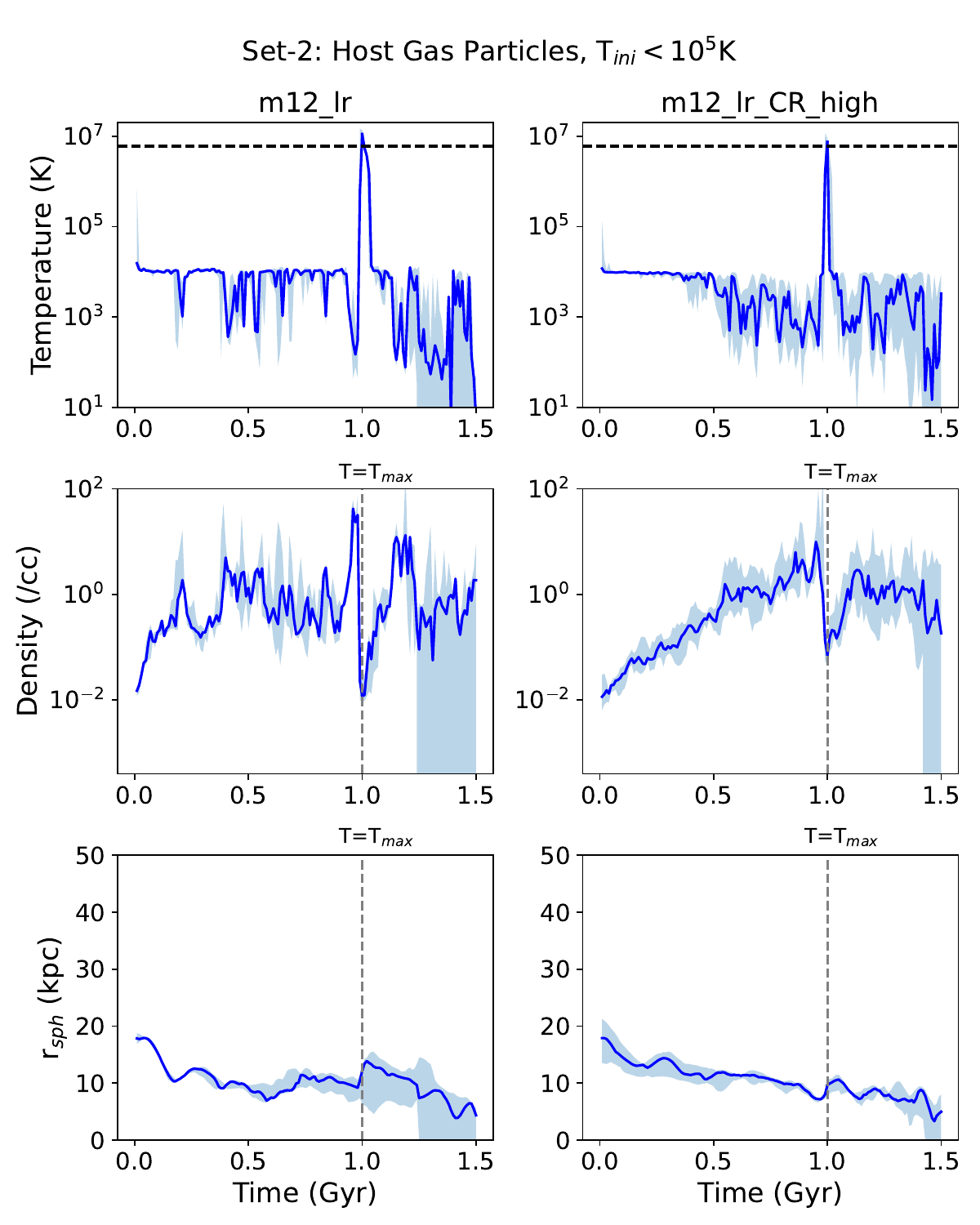}
\caption{Time evolution of the temperature (top row), density (2nd row), and spherical radius (bottom row) of the host gas particles with initial temperature T$_{\rm ini} \sim 10^4$ K, which reached their maximum temperature at a simulation time of 1 Gyr. The solid line represents the median values of the gas properties, with the shaded regions indicating the dispersion. The figure illustrates how the cold gas of the disk is heated to super-virial temperatures due to feedback heating. These gas particles are missing in no feedback runs and have a subdominant contribution to the super-virial phase in all other runs.}
\label{f:temp_sv_cold}
\end{figure}

{In summary, we find that the total super-virial gas mass lies in the range $0.2-2\times10^8$ ${\rm M_\odot}$ as the galaxy evolves. The simulations without feedback display higher super-virial gas masses, suggesting super-virial gas does not predominantly originate from feedback heating processes. Minimal disparities in the super-virial gas mass fluctuations are observed between runs with and without cosmic rays, with higher CR runs showing slightly more pronounced fluctuations. $70-90
\%$  of the total super-virial gas lies within R$_{cyl}<15$kpc and $1<z<6$ kpc, suggesting an extra-planar geometry extending up to $z\sim6$kpc scale height.}

\subsection{Where is the super-virial gas coming from? }
\label{S:origin}
Now the question is, what is the mechanism by which this gas is heated up? As noted above, CR heating and feedback heating do not seem to be responsible for heating to super-virial temperatures in our simulations. {Therefore, we track the host gas particles with initial temperature T$_{\rm ini} \sim 2 \times 10^6$ K, which reached their maximum temperature at a simulation time of 1 Gyr. We calculate the past temperature, density, and spherical radius of all such gas particles (host and satellite). {We find that three sets of gas particles contribute to the super-virial phase, according to their initial properties, with two sets coming from the host and one from the satellites. The first set comes from the host with an initial temperature similar to T $>8\times10^5$K (set-1 hereafter). The second set is also from the host with the initial temperature of a cold gas T$<10^5$K (set-2 hereafter). There is another set (set-3 hereafter), made of gas particles initially coming from satellites and found to be at super-virial temperature after $>1$ Gyr. We track these particles and plot the median and dispersion of properties of all the trajectories of these particles for each case.} We list the properties of set-1, set-2 and set-3 particles below.

\begin{figure}
\includegraphics[width=\linewidth]{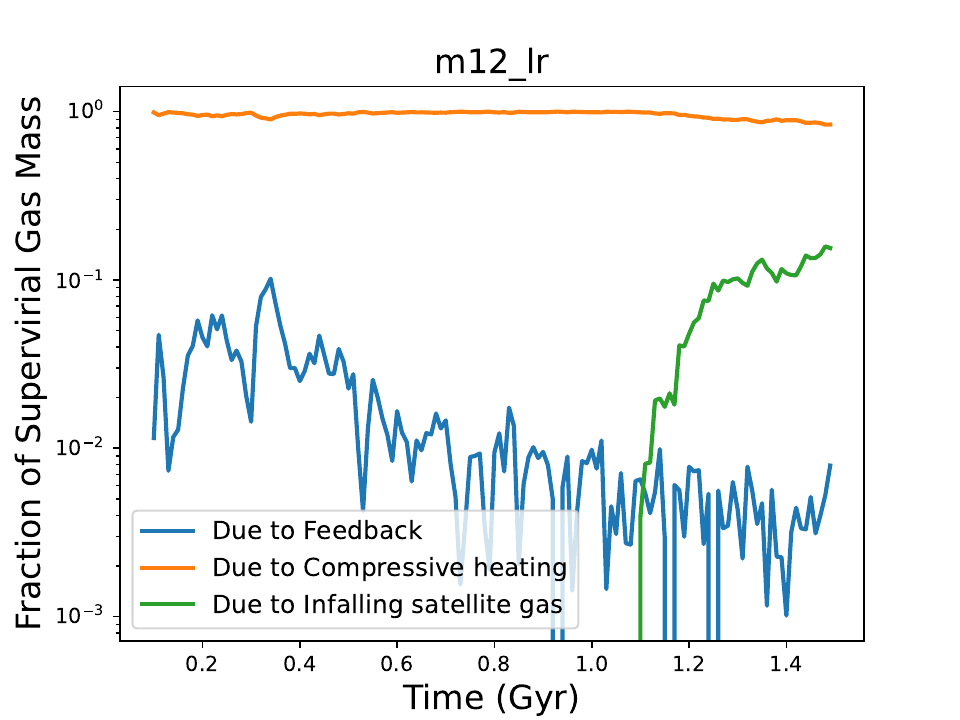}
\caption{Fraction of super-virial gas mass attributed to stellar feedback (initial temperature T$_{\rm ini} \sim10^4$ K) and compressive heating of infalling gas (initial temperature T$_{\rm ini}  \sim2\times10^6$ K) in the case of our fiducial run. The figure shows that feedback contributes approximately 1\% to the super-virial gas mass in our simulations, while infalling virial gas is the dominant contributor to the super-virial phase. Infalling satellite gas also starts to contributes 3\% at t=1.2 Gyr, increasing their contribution to 10\% at t=5 Gyr.}
\label{fraction}
\end{figure} 

\begin{figure*}
\includegraphics[width=\textwidth]{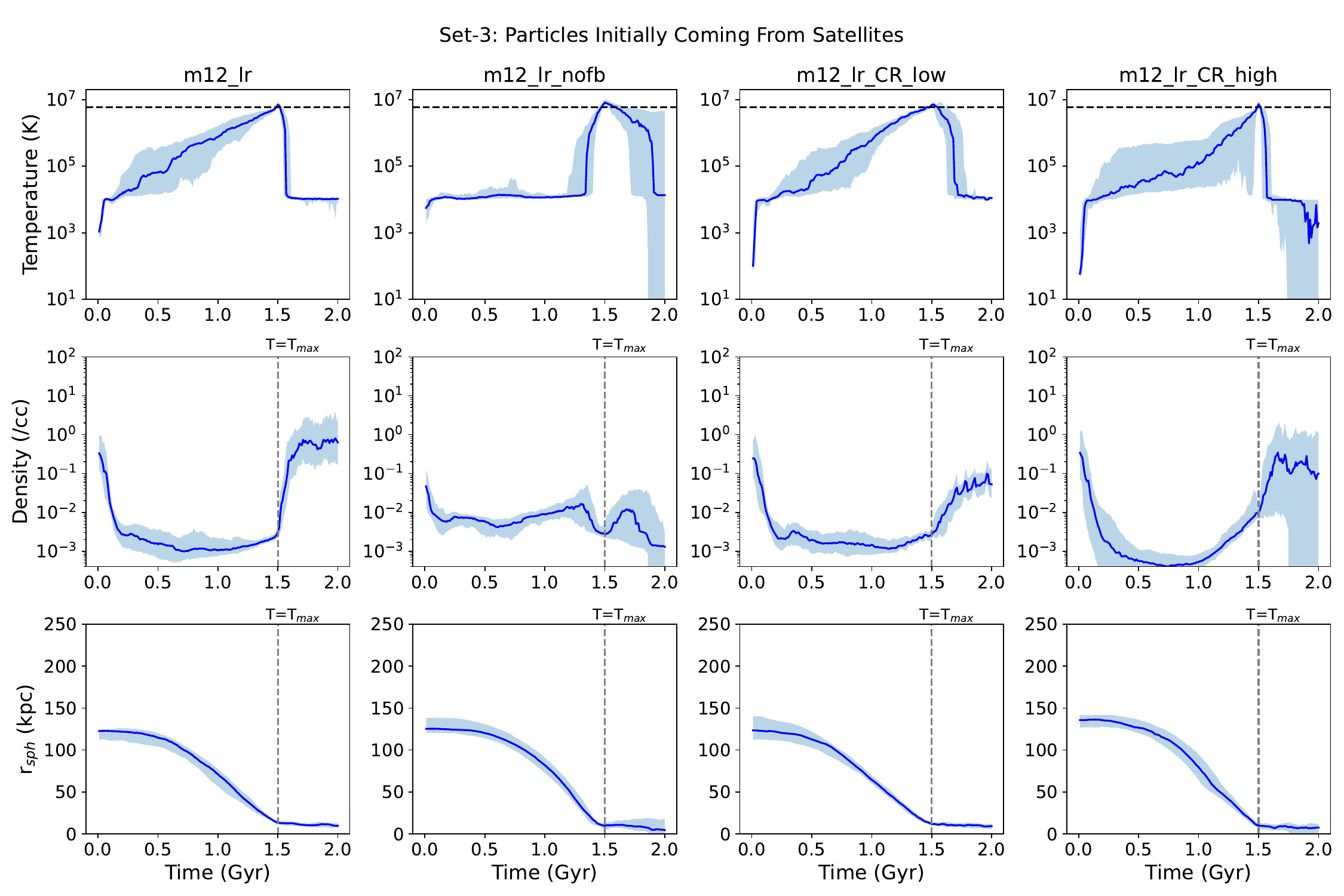}
\caption{Time evolution of the temperature (top row), density (2nd row), and spherical radius (bottom row) of gas particles initially coming from satellites, which reached their maximum temperature at a simulation time of 1.5 Gyr. The solid lines represent the median values of the gas properties, while the shaded regions indicate the dispersion. The figure illustrates how infalling gas originating from satellites is heated to super-virial temperatures through compressive heating.}
\label{f:sat_sv}
\end{figure*} 

\begin{figure*}
\includegraphics[width=0.33\textwidth]{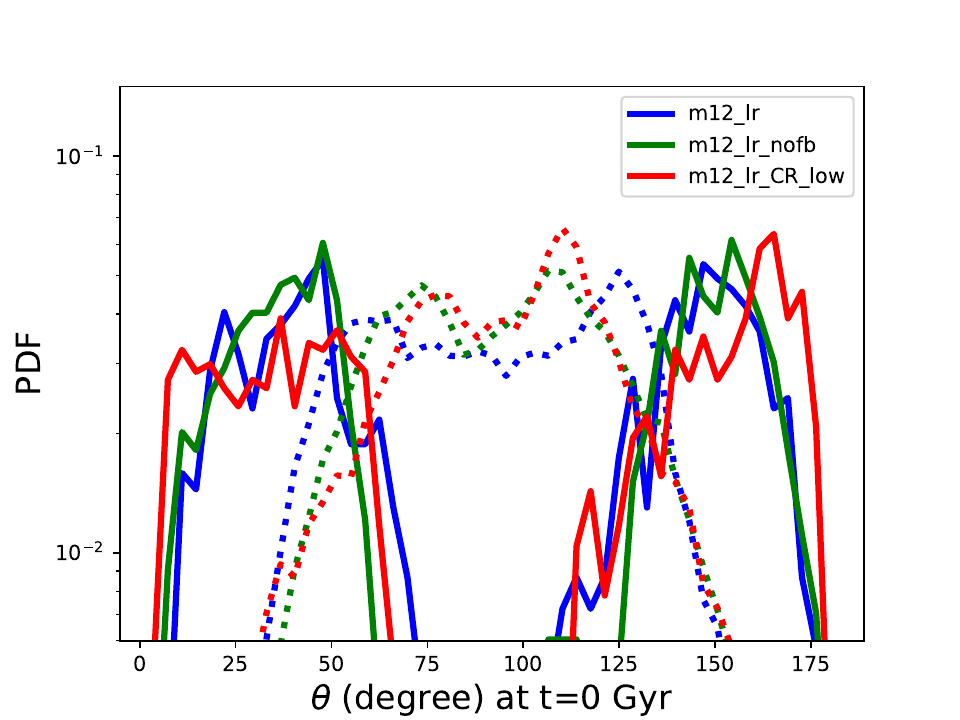}
\includegraphics[width=0.33\textwidth]{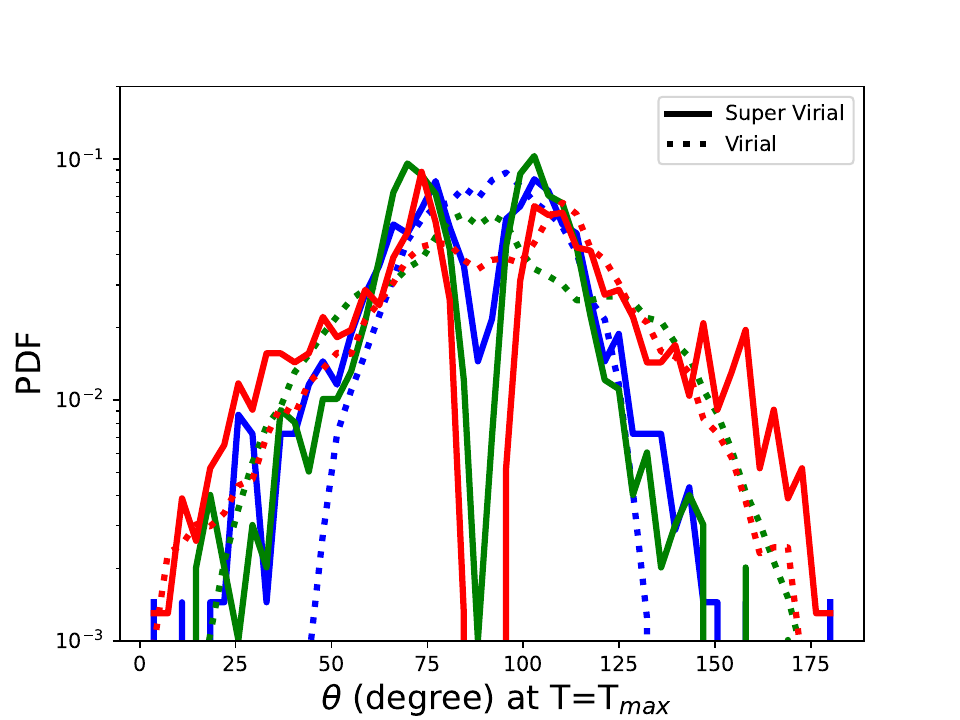}
\includegraphics[width=0.33\textwidth]{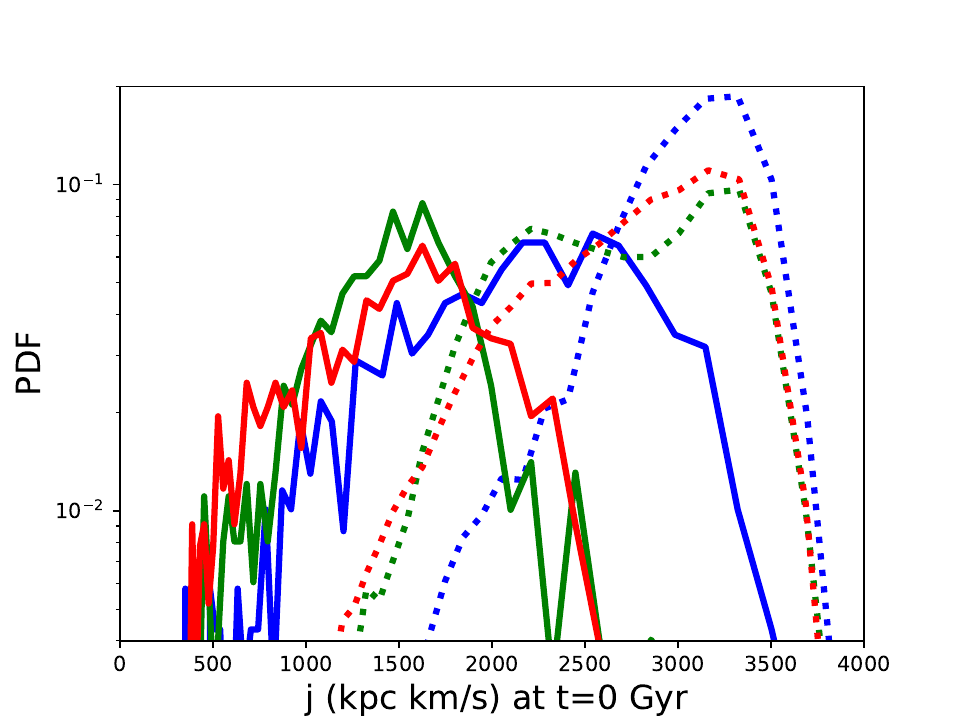}
\includegraphics[width=0.33\textwidth]{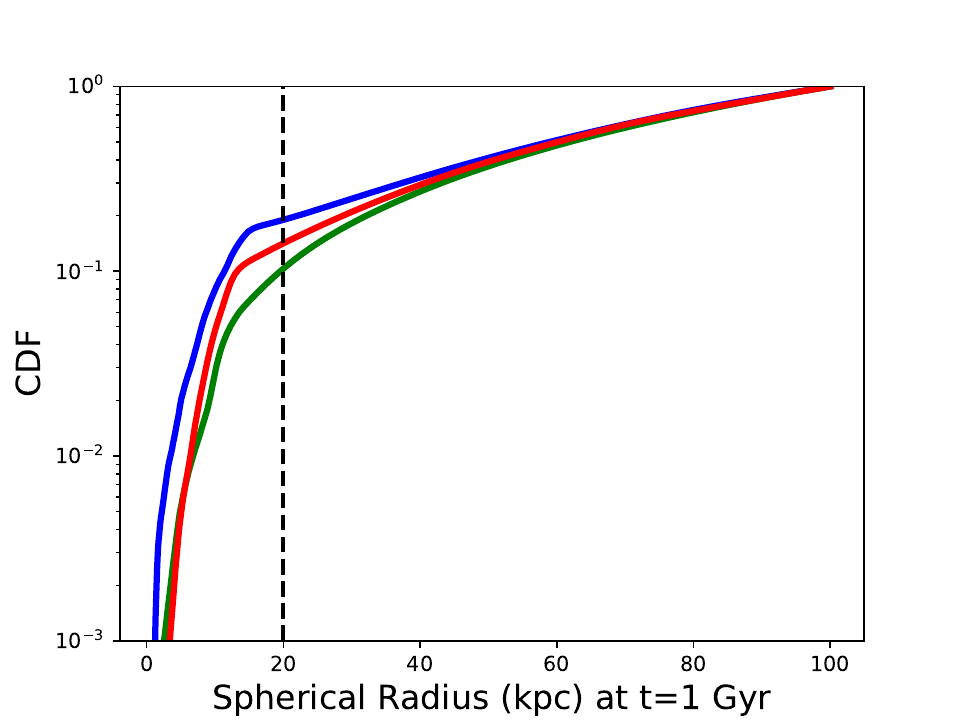}
\includegraphics[width=0.33\textwidth]{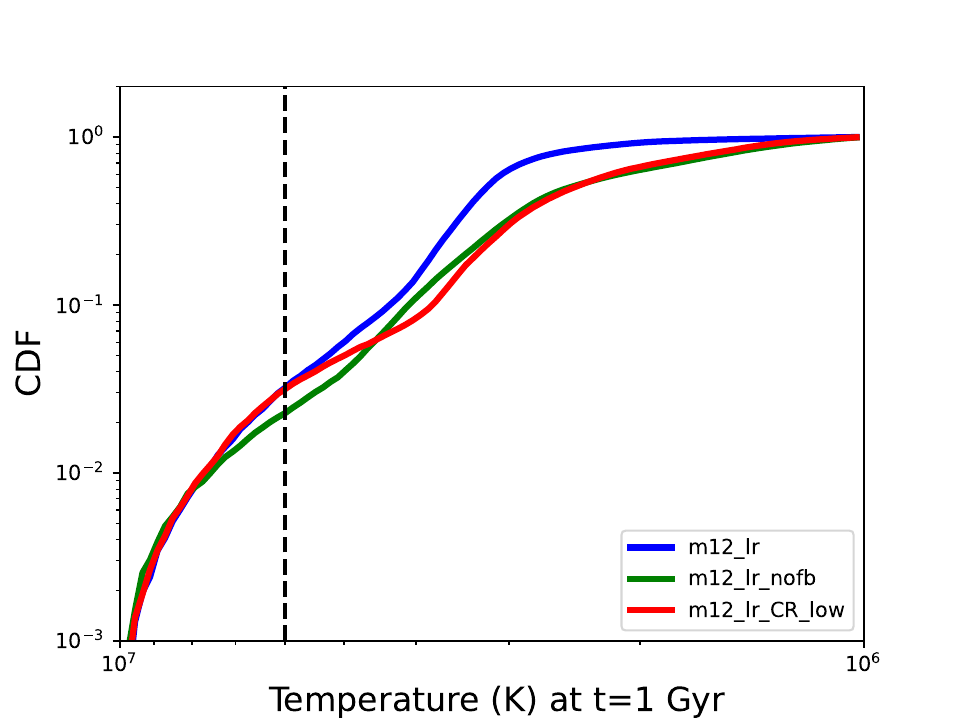}
\caption{\textit{Top:} Probability distribution function of the angle {subtended by the particle position} to the galaxy rotation axis at $t=0$ Gyr (left panel), at maximum temperature $T_{\rm max}$ (middle panel), {and specific angular momentum at $t=0$ Gyr (right panel)} for all infalling gas particles with a virial temperature beyond 20 kpc. Dotted and solid lines indicate virial and super-virial temperatures within 20 kpc at t=1 Gyr, respectively. Gas that heats up to super-virial temperatures starts closer to the rotation axis {and has less angular momentum}, while gas farther from the axis {with more angular momentum} remains at virial temperatures. The middle panel shows that when the gas reaches super-virial temperatures, it is more uniformly distributed across the disk than initially.
\textit{Bottom:} Cumulative distribution function (CDF) of the spherical radius (left panel) and temperature (middle panel) at t=1 Gyr of gas particles that start at virial temperature beyond 20 kpc at t=0 Gyr. The figure demonstrates that around 10-15\% of the gas is found within 20 kpc at $t=1$ Gyr. About 1\% of this infalling virial gas is heated up to super-virial temperatures.}
\label{temp_cdf}
\end{figure*}

\subsubsection{Set-1: Host gas particles, T$_{\rm ini}>8\times10^5$K}
In Figure \ref{f:temp_sv}, blue curves show the time evolution of the temperature (top row), density (2nd row), and spherical radius (3rd row) of host gas particles {with initial temperature T$_{\rm ini} > 8 \times 10^5$ K}, which reached their maximum temperature at a simulation time of 1 Gyr. The top panel shows that these gas particles roughly have an initial temperature $\sim2\times10^6$K in all the cases except for the high CR run, where the initial temperature was $\sim1\times10^6$K. This is because, for the high CR case, the initial halo has a lower thermal pressure and, in turn, a lower temperature halo. The gas heats up slowly over time and reaches over $6\times10^6$K at t=1 Gyr.
In the third panel of Figure \ref{f:temp_sv}, we plot the time evolution of the position of the super-virial gas particles. All of them are falling from the outer halo (50-80 kpc) into the central 10 kpc region. For the high CR run, the gas particles fall from a little farther away, $80-90$ kpc, and for all the other cases, the particles fall from 40-50 kpc. {The overall gas temperature of the CGM volume filling hot gas of the CGM in the high CR run is lower by a factor of two, as a result of our requirement to maintain HSE in the initial conditions. Therefore, the hot virial gas infalling from farther out than other runs can only heat up to the super-virial temperature, and hence, super-virial gas appears later in this run than other runs, as shown in the top-left panel of Figure \ref{f:time_evo_sv}.}

In the second row of Figure \ref{f:temp_sv}, we plot the past density of these super-virial gas particles. All of them start from $10^{-3} {\rm cm}^{-3}$, which is a typical density of the diffuse virial gas. {As their temperature increases, their density also slowly increases over time as they fall inwards}. Following their rise to super-virial temperatures at 1 Gyr, they cool down, increasing their density to $1\,\rm cm^{-3}$.
Again, the high CR run starts from a lower density of $5\times10^{-4}$$\rm cm ^{-3}$ than the density of other runs, as we are tracking the gas particles from a larger radius, as we mention in the next paragraph. Eventually, they reach the same final density as all the other runs. {Therefore, there is almost a factor of $\sim5-10$ times increase in density while heated up, whereas $\sim1000$ times increase in density after they cooled down.}

{In the fourth panel of Figure \ref{f:temp_sv}, we show the trajectory of these particles in the $R_{\rm cyl}-z$ plane. This figure shows that the particles start from higher $R_{\rm cyl}$ and z and land at smaller $R_{\rm cyl}$ and z, which is near the galactic disk.} 
 
{As the gas particles fall inwards, they start heating up as this infall changes the gravitational potential energy, which can be converted to thermal energy and slowly heat the gas. Until 0.9 Gyr, the temperature only increases to $ 3-4\times10^6$ K. At 0.9 Gyr, these gas particles roughly reach 10 kpc, which is the radius of the galactic disc. At that point, they either get compressed {by higher pressure} and heated to over $6\times10^6$ K or they can get shock-heated. Therefore, it is clear that set-1 gas particles are infalling onto the disc, and they are heated up by {adiabatic compression and/or shocks}, and hence we see enhancement of density and temperature.} Once they heat up to the super-virial phase, they cool down immediately because they become rotationally supported and radiate thermal energy away after they join the disc, as shown by the time evolution of the spherical radius $r$, which does not decline and remains constant after $t = 1$ Gyr.  {Now the question is whether it is a result of a shock or of adiabatic compression.} The initial smooth evolution of the density and temperature curves is indicative of the heating by compression rather than shocks, as shown by tracking of individual particles in Figure \ref{f:temp_sv}. {To investigate this further, we track the radial velocity of these particles over time in the 5th panel of Figure \ref{f:temp_sv}. All the runs show fluctuation in velocity between super-sonic velocity $v_{sound, T_{hot}}\sim100$ km/sec and sub-sonic velocity. This implies that it is most probably a combination of shock (where super-sonic) and adiabatic compression (where sub-sonic). Additionally, the 6th row of Figure \ref{f:temp_sv} shows the tracking of entropy for the hot inflowing particles that heat up to super-virial at 1 Gyr. It shows a roughly constant entropy throughout, with an entropy jump at 1 Gyr, where heating happens. Therefore, it implies that the earlier smooth temperature increase is due to the adiabatic compression of virial gas from infalling into a denser part of the galaxy. However, at a later time, when the entropy jump occurs, part of the heating originates from shock. Therefore, it is the combination of shock and adiabatic compression that heats the gas to super-virial temperatures.} In conclusion, the infalling gas predominantly produces super-virial gas in our simulation. Hence, we also see more super-virial gas in the no feedback run as the feedback exits along the vertical axis of the galaxy and directly opposes the infalling gas. As mentioned before, our ``no-feedback run'' is performed mainly as a control experiment to see the effect of a pure cooling flow. It enables us to disentangle the net effect of the cooling flow from that of the stellar feedback in creating the super-virial gas. Also, the CR feedback can put additional resistance to the infalling gas, creating more fluctuations. 

{In the bottom-left panel of Figure \ref{f:time_evo_sv}, we show the time evolution of the net mass flux of the hot gas at 10 kpc. We find that the value of net mass influx of the hot gas is $\sim5-10\, M_\odot$/yr, negative signs denoting inflow. However, the high-CR run shows a stronger inflow at a later time due to its lower CGM temperature. {The inflow rate is roughly comparable to the average star formation rate (SFR) of the host galaxy, shown in the bottom-right panel of Figure \ref{f:time_evo_sv}. In simulations including stellar feedback, the system evolves toward a self-regulated quasi-steady state, where the SFR becomes comparable to the hot gas inflow rate. The typical SFRs lie in the range that is consistent with observed values in Milky Way-mass galaxies.}}

{We also plot the subset of the trajectories of these particles in Figure \ref{f_a:temp_sv} to show that the median and dispersion capture the essence of the trajectories well. When we look at the individual particle trajectories, we see that most of the particles follow the above theory. In some cases, however, the particles are located near the disk, and they have cooled down. They are later lifted by feedback and heated to super-virial gas. Thus, the stellar feedback contributes to the super-virial phase for set-1 particles, but only marginally ( $<1\%$ by mass).}


\subsubsection{Set-2: Host gas particle, T$_{\rm ini}<10^5$K}
In the top row of Figure \ref{f:temp_sv_cold}, we plot the time evolution of the temperature (top row), density (2nd row), and spherical radius (bottom row) of the host gas particles {with initial temperature T$_{\rm ini} < 10^5$ K}, which reached their maximum temperature at a simulation time of 1 Gyr. The gas particles in this set are very few, except for the high CR case, and are absent in the no feedback case. These gas particles get colder and hotter over time by fluctuating between $10^4$ K and $10$K until 0.9 Gyr, where the gas heats up and reaches over $6\times10^6$K at about 1 Gyr. These gas particles are heated from stellar feedback and hence are absent in the no feedback case, and are more in the high CR case due to the additional heating from CR feedback by a larger amount of CR. 

In the middle row of Figure \ref{f:temp_sv_cold}, we plot the past density of these gas particles. All of them start from $10^{-2}$$\rm cm ^{-3}$ and finally become $10^{2}$$\rm cm ^{-3}$ right before the heating up at 1 Gyr. Then their density again drops down to $10^{-2}$$\rm cm ^{-3}$ once they are heated up by feedback. In the bottom row of Figure \ref{f:temp_sv_cold}, we plot the past position of this super-virial gas. All of them remain in the 10 kpc to 20 kpc region. 
   
Thus, we see that these set-2 gas particles represent the cold ISM gas, which gets colder and denser over time and may form stars. {Then they suddenly heat up by stellar feedback. However, set-2 gas particles are less than 0.01 times the set-1 gas particles. As we can see in Figure \ref{fraction}, the fraction of total super-virial gas host mass is $\sim1\%$ due to feedback and $\sim99\%$ due to compressive heating. This implies the mass in the super-virial phase, as traced by set-2 gas particles, is smaller by a factor of $\sim100$; hence, set-2 gas particles contribute very little to the super-virial gas mass budget.}

\subsubsection{Set-3: Particles Initially Coming From Satellites}
There is another set of gas particles that can heat up to super-virial temperature, as shown in Figure \ref{f:sat_sv}. The gas from satellites gets stripped and falls inward. As they accrete onto the galactic disc, they heat up due to the gravitational potential energy loss and compression, just as the  set-1 gas particles.
We showed in Figure \ref{fraction} that the fraction in this set is initially zero but starts developing at the timescale of satellite gas infall, around 1.2 Gyr. It starts with as low as 0.1\% of the super-virial budget but slowly goes up to even 10\% at 1.5 Gyr. Therefore, at a later time, the contribution from the satellite gas phase can also contribute significantly to the super-virial gas phase. 
\subsection{What distinguishes gas which heats up to super-virial temperatures?}
 
Orange curves in Figure \ref{f:temp_sv} plot the time evolution of the temperature (top row), density (2nd row), and spherical radius (3rd row) of infalling virial gas particles that do not heat up to super-virial temperatures. As we can see, they are also falling inwards, and as they do that, the temperature and density of the gas increase slightly.
{We can see that the gas, which is heated to super-virial temperatures, settles to the disk at smaller radii than the gas that does not reach super-virial temperatures. Smaller radii imply higher densities, which suggests more compression. In the 4th panel of Figure \ref{f_a:temp_sv}, we also see that the gas reaching the super-virial phase is closer to the rotation axis than the gas that fails to reach super-virial temperatures. We dig into this issue further and calculate the angle subtended by the particle position with the rotation axis ($\theta$) for these gas particles.}

In the top panel of Figure \ref{temp_cdf}, we plot the probability distribution function of the angle ($\theta$) {subtended by the particle position} to the galaxy rotation axis at t = 0 Gyr (left panel) and at maximum temperature (middle panel), for all infalling gas particles with a virial temperature beyond 20 kpc. In the Left panel, we can see that the gas particles that heat up to super-virial temperatures are closer to the rotation axis, whereas the gas particles that remain at virial temperature are further from the rotation axis. It implies that infalling virial gas, which originates from close to the rotation axis, reaches higher temperatures. This is consistent with the behavior of rotating hot inflows (or `rotating cooling flows') where the hot CGM phase inflows and spins up on a cooling timescale \citep{Hafen2022, Stern2024}. In these accretion flows, gas has lower densities near the rotation axis due to rotation support (see figure 4f in \citealt{Stern2024}). When the hot inflow approaches the disk, it transitions from a spherical geometry to a disk geometry. During this change in geometry, gas particles near the rotation axis suffer smaller radiative losses (which scale as density squared, see figure 4k there) and experience larger compressive heating rates (see figure 4l there). 

Supporting the rotating hot inflow interpretation, the middle panel of Figure \ref{temp_cdf} shows that the $\theta$ distribution, when the temperature is maximal, is close to 90 degrees, distinctly different from the initial $\theta$ distribution. This means that the super-virial gas particles have indeed transitioned from a spherical to a disk geometry. Further support for this interpretation is that the gas cools immediately after reaching super-virial temperatures (Figure \ref{f:temp_sv}), as expected in rotating hot inflows, which cool immediately onto the ISM after transition into a disk geometry \citep{Stern2024}. 

{In the right panel of Figure \ref{temp_cdf}, we also present the probability distribution function of specific angular momentum j at t = 0 Gyr. We see that lower angular momentum gas that is near the rotation axis is forced to fall directly down the axis and heats up. On the other hand, gas further from the rotation axis with a larger j does not fall directly and is not heated up to high temperatures. Therefore, the origin of the gas heated to super-virial temperatures is largely influenced by our axisymmetric model, where the halo rotates around a central axis.}


In summary, some of the virial temperature gas is infalling from the outer CGM ($50-60$\, kpc) to the inner CGM ($20-10$\, kpc). All of this inflowing gas loses gravitational energy and is heated up by a factor of $1.5-2$. 
When it reaches the inner CGM, the gas gets further heated up by another factor of $1.5-2$ due to the combination of compressive heating and shock heating when the hot inflow changes from a spherical to a disc geometry. This change in geometry is instigated by rotation support in the hot inflow. 
Immediately thereafter, the hot inflow cools and joins the ISM disc. 
Specifically, inflowing virial gas closer to the rotation axis of the galaxy suffers less radiative cooling and more compressive heating and shock heating due to lower gas density near the axis and, therefore, heats more.  

\subsection{What amount of infalling virial gas heats up to the super-virial phase?}
To investigate how much virial gas is infalling, we calculate a cumulative distribution function (CDF) of the final position (at t=1 Gyr) of the gas particles, which starts at virial temperature beyond 20 kpc at t=0 Gyr. In the left panel of the bottom row of Figure \ref{temp_cdf}, we show that only 10\% of gas particles enter 20 kpc at $t=1$ Gyr. With these gas particles, which are 20 kpc at t=1 Gyr, we make a CDF of temperature distribution in the middle panel of the bottom row of Figure \ref{temp_cdf}, where we see only {2-3\%} of these gas will heat up to super-virial temperatures.

\subsection{Fate of this super-virial gas}
The super-virial gas will also eventually cool down once it is supported by rotation, as we see in  Figure \ref{f:temp_sv}, \ref{f:temp_sv_cold}, and \ref{f:sat_sv}. We can see all the super-virial gas cools down in the later snapshot after they reach the maximum temperature, consistent with the hot rotating inflow solution discussed in \cite{Stern2024}. But the cooled gas will immediately be replenished by heating the continuously inflowing virial gas by the same process. Therefore, we see a constant amount of super-virial gas over time.

To summarize, we investigate mechanisms driving the heating of the virial gas within a galaxy, finding that gas particles exhibit distinct heating behaviors based on their initial conditions and interactions. Set-1 gas particles, originating from the halo, closer to the rotation axis, heat up as they infall, where the gas converts gravitational energy to heat mainly {via a combination of compressive heating and shock heating}. This process causes gas falling close to the rotation axis to reach super-virial temperatures just before cooling and joining the disk. In contrast, set-2 gas particles, likely representing the cold ISM gas, heat up sporadically, potentially due to stellar feedback, contributing minimally to the super-virial gas mass budget. Additionally, gas particles stripped from satellites and accreting onto the disc also heat up via compressive heating, further contributing to the super-virial phase. Some infalling virial gas particles fail to heat up to super-virial temperature as they fall further from the rotation axis, hence suffering more radiative cooling and less compressive heating.

\section{Discussion, Limitations and Future work} \label{dis}
We also investigate the temperature profile in a run where the virial temperature phase of the CGM is initialized with a cooling flow rather than hydrostatic equilibrium, using the cooling flow solution from \cite{Stern2019}. We find that our cooling flow run also produces super-virial gas within 20 kpc, and its mass is of the same order as the other runs (Figure \ref{f:time_evo_sv}). This makes sense since HSE initial conditions are expected to develop into a cooling flow within a cooling time \citep{Stern2019}. 

The top panel of Figure \ref{f:2d-pdf} indicates that the super-virial gas is extra-planar, just above the ISM disk. It will thus be very interesting to investigate future X-ray observations of edge-on galaxies, which we will discuss in the follow-up paper. In future work, we will also post-process and produce observables like X-ray emission and absorption maps from the simulation snapshots. This will both facilitate direct comparison with observed maps and assist in distinguishing between emission and absorption origins. 

\textbf{Effect of different feedback:} It is fascinating to see that conventional CR/feedback heating does not produce this super-virial phase as anticipated, whereas gravitational and compressive heating is the dominant heating mechanism in our idealized {\small GIZMO} simulations with the FIRE stellar feedback model. However, this still depends on the feedback prescription used in the simulation. With a different feedback prescription \citep[e.g.,][]{2024MNRAS.527.1216S,2024arXiv240815321S}, one may get a different result. For example, the idealized simulation by the recent work of \cite{2022MNRAS.510..568V} reproduced this temperature component, likely caused due to heating by stellar feedback, but with a very high star-formation rate, unlike the MW. Another recent work by \cite{Bisht2024} concludes that the super-virial gas results from the stellar feedback in their idealized simulation. Additionally, we do not have AGN feedback, which can have a larger impact than stellar feedback in terms of heating the gas \citep[e.g.,][]{2023MNRAS.523.1104W,2023MNRAS.523.2409C,2023arXiv231214809Q,2024MNRAS.532.2724S}. For example, recent work by \cite{2023MNRAS.518.5754R} also examines the physical properties of gas within the CGM surrounding 132 MW-like central galaxies at a redshift of z=0, using data from the cosmological simulation TNG50, which is part of the IllustrisTNG project. They found that energy from the supermassive black hole (SMBH)-driven kinetic winds heats the gas to super-virial temperatures (exceeding $10^{6.5}$ to $10^7$ K). Therefore, it would be useful to explore several different simulations and investigate similar questions to compare. It will not only give us an overall picture of this newly found super-viral phase, but also constrain varying feedback mechanisms in different simulations. 

\textbf{Absence of cosmological context:} Another limitation of our work is that it does not consider the galaxy in the cosmological context. {The CGM of real galaxies is shaped by a complex cosmological history, involving non-steady accretion, feedback-driven baryon cycling, and chemical inhomogeneities. While our simulations do not include full cosmological evolution or cold-mode filamentary accretion, they are purposefully designed to isolate and study the behavior of hot-mode accretion in a controlled environment. Our initial conditions feature a simulation domain that extends to three times the virial radius, providing a natural reservoir of hot gas that can flow inward during the simulation. This setup captures gravitational infall of gas at roughly the virial temperature into the inner halo, our region of interest, over the timescale of the simulation. Importantly, our focus is not on cosmological accretion per se, but rather on how virial-temperature CGM gas infalls toward the galactic disk and gives rise to the super-virial phase observed in X-ray studies. We emphasize that we are not claiming this is the only source of the super-virial gas, but are testing it as one of the sources. We also argue that a steady cooling flow, although simplified, can describe the zeroth-order effect of the hot accreting gas in cosmological simulations (e.g., \citealt{2024arXiv241016359S}). While the absence of cold filaments and cosmological assembly history is a caveat, our goal is to disentangle the hot gas dynamics in isolation, which is difficult to achieve in fully cosmological simulations due to their inherent complexity.} 

{In addition, our simulations adopt a simplified, axisymmetric, and isolated galaxy model to enable a controlled investigation of hot inflow and the resulting formation of super-virial gas. While this setup allows us to disentangle specific physical processes, it does come with limitations. In particular, our CGM is initialized as a smooth, hydrostatic, and baryon-rich medium. {Despite uncertainties remaining in the baryonic fraction of Milky Way-mass galaxies, it may have a higher baryon fraction than the Milky Way \citep{Miller2013, Miller2015} due to the absence of prior baryon loss from feedback or cosmological assembly.} As this CGM cools, it naturally begins to collapse and drive inflows, with the infall rate evolving non-monotonically over time. This {idealized} setup may enhance infall and subsequent heating, potentially exaggerating the generation of super-virial gas in the early phases. 
{As for the axial symmetry, while the net angular momentum assumed in our initial conditions is broadly consistent with full cosmological simulations \citep{De2020} and should capture the zeroth-order inflowing angular momentum of hot accreting gas, it may also result in a more coherent inflow of gas than occurs in reality.}
While we find qualitative similarities with expectations from cosmological simulations, such as feedback-driven regulation and transient heating, we acknowledge that the detailed evolution is sensitive to the initial conditions and timing of measurement. A fully cosmological context will be necessary to test the robustness of these results, which we leave for future work.} Our future works will explore the same questions with different cosmological simulations, such as FIRE zoom-in runs \citep{2017arXiv170206148H,2023MNRAS.519.3154H,2023MNRAS.520.5394W}, HESITA \citep{HESITA2020}, TNG \citep{TNG1,TNG2}, and EAGLE \citep{EAGLE1, EAGLE2} simulations. These insights will deepen our understanding of galactic gas dynamics and provide valuable implications for broader inquiries into galaxy evolution and structure formation. 

{\textbf{Effect of ad-hoc satellite distribution:} The satellite configuration was originally designed to see satellite contribution to the super-virial gas, and it was found to be insignificant. However, such exploration is still important since we covered a wide parameter space of satellite mass and orbit, and still did not see a huge effect. The timescale of $\sim1-1.5$ Gyr, when satellites start to contribute to super-virial gas, is indeed a product of the initial conditions. The infall time scale of these satellites from 100-150 kpc is roughly $\sim1-1.5$ Gyr. A different satellite distribution in initial conditions can potentially make their effect important according to their infall time.} 

{In summary, while our idealized simulations lack a full cosmological context, they provide a valuable framework for isolating and understanding the physical processes that give rise to super-virial gas in a controlled environment. This approach allows us to disentangle the roles of hot inflow, feedback, and gravitational dynamics without the added complexity of ongoing structure formation. Additionally, our exploration of satellite contributions sets the stage for future work using cosmological simulations, where such interactions will occur more naturally. Finally, this setup offers a useful testbed for examining the efficacy of FIRE stellar feedback in driving thermal evolution of the CGM; notably, we find that the feedback alone does not appear sufficient to generate significant super-virial heating in the absence of additional infall-driven processes.}

\section{Conclusions}
\label{S:conclusion}
Several observations have recently established that the MW contains gas at temperatures higher than its virial temperature. However, the location, origin, and properties of this gas are still unknown. In this paper, we analyze a suite of idealized simulations to investigate this newly-discovered gas phase and find this super-virial phase gas in our simulations of a {MW-mass galaxy}. Our main findings are the following:

\begin{itemize}
\item Most of the super-virial gas is within 20 kpc from the center of the galaxy. {70--90\% of the total super-virial gas is at $1<z<6$ kpc and $R_{\rm cyl}<15$ kpc (top panel of Figure \ref{f:2d-pdf} and Figure \ref{f:time_evo_sv}). This implies that it is situated mostly in the extra-planar disc region, not in the extended CGM.}

\item This super-virial gas phase is found close to the disk where typical gas densities are {$10^{-3}-10^{-2}\, {\rm cm}^{-3}$} (Bottom Panel of Figure \ref{f:2d-pdf} and 2nd panel of Figure \ref{f:temp_sv}, and bottom panel \ref{f:temp_sv_cold}).

\item The mass of this gas phase is roughly constant over time, in the range $0.2-1\times10^8\,{\rm M_\odot}$, {but does not significantly contribute to the CGM mass budget}. This phase is retained in the simulations even at 1.5 Gyr (Figure \ref{f:time_evo_sv}).

\item We also investigate the effect of stellar feedback in heating the gas. 
The feedback heating contributes only less than 1\% of the super-virial gas, implying the mechanism of feedback heating is sub-dominant, at least for the FIRE-feedback prescription (Figure \ref{fraction}). 

\item We also explore the effect of CR in the heating of the gas. With high and low CR initial setups, we found minimal differences in the super virial gas mass budget, however, the gas mass in CR runs shows more fluctuation, caused by additional CR feedback effect. Therefore, it implies that the CR has little role in heating the gas to super-virial temperatures (Figure \ref{f:time_evo_sv}).

\item While investigating the origin of super-virial gas, we found that the dominant contribution to heating is caused by the adiabatic compression and shocks of the rotating inflow of the virial CGM phase, as in the model of \cite{Stern2024}. In this model, gravitational energy is converted to heat mainly via compressive heating in the inflow. {Later on, they are shock-heated before settling onto the disc, indicated by the entropy jump (bottom panel of Figure \ref{f:temp_sv}).} We show this process causes the {$2-3$\%} of infalling virial gas, closest to the rotation axis, to reach super-virial temperatures just before cooling and joining the disk (Figures \ref{temp_cdf}). Gas inflowing near the rotation axis has a lower density and thus suffers less cooling losses and is more compressively heated than gas farther away from the axis, allowing it to reach higher temperatures. Gas stripped from satellites also inflows onto the central galaxy and heats up similarly (Figure \ref{f:sat_sv}).


\end{itemize}


{In conclusion, we identify a combination of compressive heating and shock-heating of the hot inflow as a significant contributor to the super-virial gas phase in our MW-mass galaxy simulation using FIRE-2 physics. However, this process is not the only channel, but rather one of the several channels by which super-virial gas can be formed. Other channels include feedback heating and CR heating, which are not dominant in our simulation and are explored in other studies. In reality,  the combination of all of these processes can give rise to the super-virial phase in a galaxy.} 

\newpage
\section*{Acknowledgments}
We thank Todd Thompson and the FIRE collaboration for useful discussions and suggestions. MR acknowledges support from the CCAPP fellowship at The Ohio State University and ACCESS allocations of PHY240003. MR acknowledges the Aspen Center for Physics and Simons Foundation, as part of this work was performed there, which is supported by a grant from the Simons Foundation (1161654, Troyer). KS acknowledges support from the Black Hole Initiative at Harvard University, which is funded by grants from the John Templeton Foundation and the Gordon and Betty Moore Foundation and acknowledges ACCESS allocations TG-PHY220027 and  TG-PHY220047 and Frontera allocation AST22010. SM is grateful for the grant provided by the National Aeronautics and Space Administration (NASA) through Chandra Award Number AR0-23014X issued by the Chandra X-ray Center, which is operated by the Smithsonian Astrophysical Observatory for and on behalf of the National Aeronautics Space Administration under contract NAS8-03060. S.M. is grateful for the NASA ADAP grant 80NSSC22K1121. JS was supported by the Israel Science Foundation (grant No.\ 2584/21).  The computations in this work were run at facilities supported by the Bridges, University of Pittsburgh.

\appendix
\setcounter{figure}{0}  
\renewcommand{\thefigure}{\thesection.\arabic{figure}}
\section{Additional Figures}
We show the subset of individual particle tracking for set-1 particles in Figure \ref{f_a:temp_sv}, for which we show the median with percentile error in the main text. We note that most of the particles behave similarly like the median trajectory within the error bar. Therefore, we think that the median with error is pretty representative of the whole sample. It is also true for the set-2 and set-3 particles, which we don't show here.  

\begin{figure*}
\includegraphics[width=\linewidth]{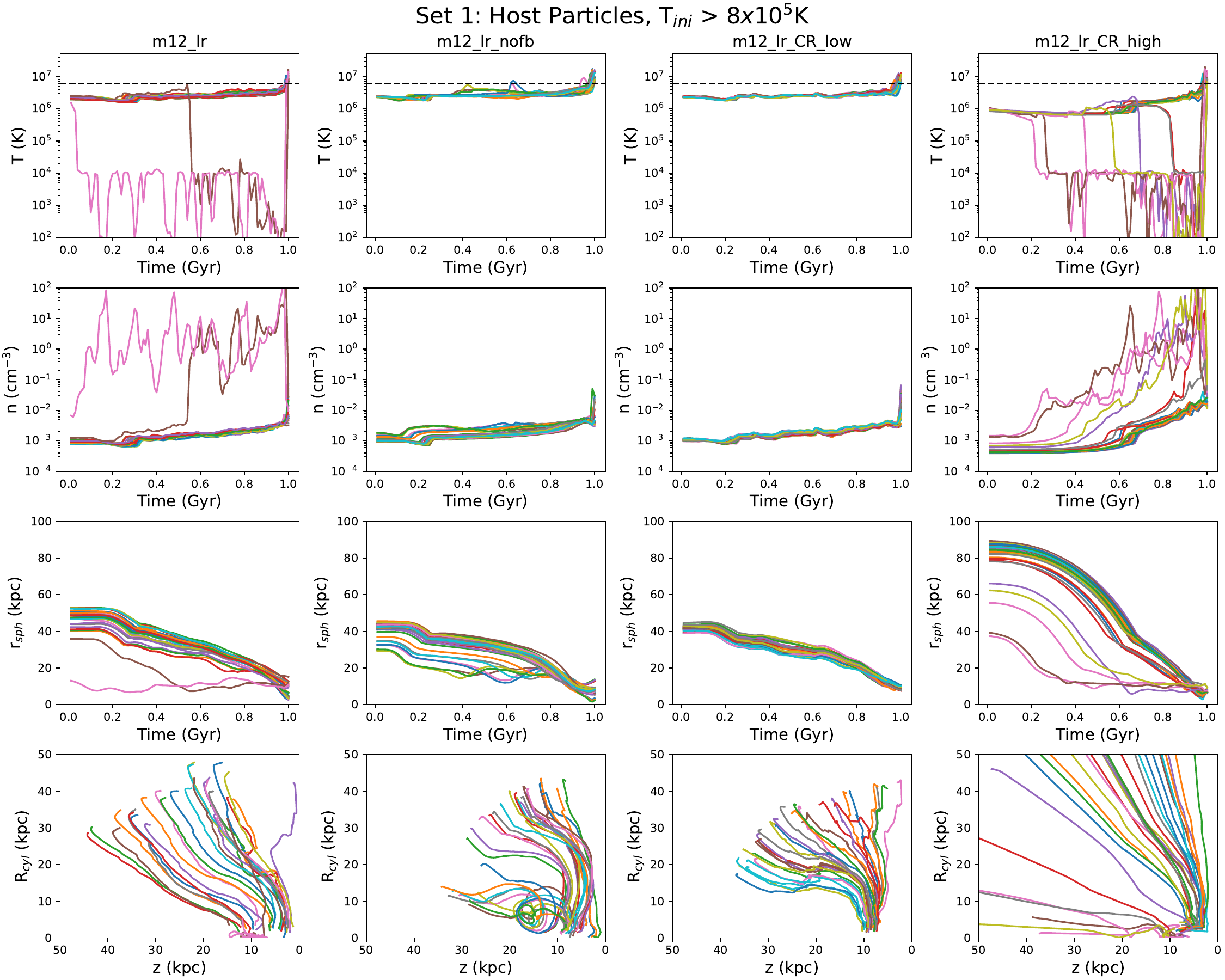}
\caption{Individual trajectories of the subset of set 1 particles shown in Figure \ref{f:temp_sv}.}
\label{f_a:temp_sv}
\end{figure*}


\normalsize

\label{lastpage}
\bibliography{main}
\bibliographystyle{aasjournal}

\end{document}